\documentclass[11pt]{article}
\usepackage[utf8]{inputenc}
\usepackage[T1]{fontenc}
\usepackage{amsmath}
\usepackage{amsfonts}
\usepackage{amssymb}
\usepackage[version=4]{mhchem}
\usepackage{float}
\usepackage{url}
\usepackage{multirow}
\usepackage{stmaryrd}
\usepackage{graphicx} 
\usepackage{booktabs} 
\usepackage[a4paper, margin=1in]{geometry}
\usepackage{tikz}
\usepackage{quantikz} 
\usepackage[authoryear]{natbib}
\usepackage{amsthm}
\theoremstyle{definition} 
\newtheorem{example}{Example}[section]

\title{Quantum Framework for Wavelet Shrinkage}
\author{Brani Vidakovic}
\date{Department of Statistics, Texas A\&M University, College Station, TX, USA}

\begin{document}

\maketitle

\begin{abstract}
This paper develops a unified framework for \emph{quantum wavelet shrinkage}, extending classical denoising ideas into the quantum domain. Shrinkage is interpreted as a completely positive trace preserving process, so attenuation of coefficients is carried out through controlled decoherence rather than nonlinear thresholding. Phase damping and ancilla driven constructions realize this behavior coherently and show that statistical adaptivity and quantum unitarity can be combined within a single circuit model. The same physical mechanisms that reduce quantum coherence, such as dephasing and amplitude damping, are repurposed as programmable resources for noise suppression. Practical demonstrations implemented with Qiskit illustrate how circuits and channels emulate coefficientwise attenuation, and all examples are provided as Jupyter notebooks in the companion GitHub repository. Encoding schemes for amplitude, phase, and hybrid representations are examined in relation to transform coherence and measurement feasibility, and realizations suited to current noisy intermediate scale quantum devices are discussed. The work provides a conceptual and experimental link between wavelet based statistical inference and quantum information processing, and shows how engineered decoherence can act as an operational surrogate for classical shrinkage.
\end{abstract}

\section{Introduction}
\label{sec:intro}

Wavelet shrinkage \citep{donoho1994,donoho1995, Nason1996, Chipman1997, Vidakovic1998} is a cornerstone of nonparametric signal estimation. A noisy signal is expanded in an orthogonal wavelet basis, where the coefficients represent local fluctuations across scales. Most coefficients are dominated by noise, while a few capture essential structural information. By attenuating small coefficients and retaining the large ones, shrinkage rules yield adaptive estimators that preserve edges and discontinuities while achieving near-minimax performance across broad smoothness classes. This balance between localization, adaptivity, and statistical optimality has made wavelet shrinkage a central paradigm in statistical signal processing.

Quantum computation provides a fundamentally new substrate for such operations. In this setting, data are encoded not as classical vectors but as quantum states. Under amplitude encoding, a signal $x \in \mathbb{R}^N$ is mapped to the normalized state
\begin{eqnarray}
|x\rangle = \frac{1}{\|x\|}\sum_{j=0}^{N-1} x_j |j\rangle,
\end{eqnarray}
and an orthogonal wavelet transform $W$ becomes a unitary operator $U_W$ satisfying
\begin{eqnarray}
U_W |x\rangle = |W x\rangle.
\end{eqnarray}
The entire multiscale decomposition is thus executed coherently in superposition. The challenge is that denoising, traditionally a nonlinear and dissipative process, must now be expressed through linear and norm-preserving quantum operations.

Quantum analogues of the wavelet transform were first proposed in the late 1990s, when \citet{fijany1998springer,fijany1998} and \citet{Klappenecker1999}   demonstrated that orthogonal wavelet transforms can be implemented as unitary circuits acting on quantum registers. Other, more recent, references include \citet{GosalLawton2001,PuschelMouraFijany2001, DasDattaFijany2002}, to list just a few. These studies established that the hierarchical structure of classical discrete wavelet transforms admits efficient quantum realizations with complexity scaling comparable to that of the Quantum Fourier Transform. Subsequent developments generalized these constructions to multi-level and multi-dimensional cases~\citep{LiWang2018, LiLi2018, li2019}, revealing that the recursive subband decomposition inherent in wavelet analysis can be encoded as a sequence of controlled unitaries acting on amplitude-encoded data. The most recent synthesis by \citet{Bagherimehrab2023} provides a unifying framework for all quantum wavelet transforms via the linear combination of unitaries (LCU) technique, enabling systematic construction of packet-like and lifting-scheme wavelets within a quantum algorithmic setting.

Applications of QWTs now extend beyond transform synthesis. \citet{GarciaMata2009} employed the quantum wavelet transform to compute multifractal exponents of quantum states, illustrating its potential for analyzing structural properties of wave functions. \citet{Zhang2019} used QWTs for dimension reduction, where the wavelet basis served as a multiresolution subspace projection on amplitude-encoded vectors. The low-frequency subband was retained as a compressed representation, effectively halving the Hilbert-space support per decomposition level while preserving most of the signal energy. Further advances include a lifting-based quantum integer wavelet transform by \citet{ChaurraGutierrez2023} and the implementation of quantum wavelet-based denoising on physical platforms by \citet{Ma2024}, both signaling the transition of QWTs from theoretical constructs to executable quantum routines.
While these works establish the theoretical and algorithmic basis of quantum wavelet transforms, none addresses how nonlinear shrinkage or denoising could be realized within a physically consistent quantum framework. The Stinespring dilation results, in our context eauivalent to Kraus representations~\citep{stinespring1955,kraus1971,choi1975,  lindblad1976} provide a natural language for incorporating controlled decoherence or amplitude damping as a surrogate for thresholding, and ancilla-assisted operations offer a mechanism to implement nonlinear attenuation through completely positive trace-preserving (CPTP) maps \citep{Geller2024}. Yet, the existing literature treats the QWT as a strictly linear unitary transform, with no integration of channel-based or ancilla-driven shrinkage models. The absence of such nonunitary extensions defines an open frontier in wavelet based quantum signal analysis.

The present work develops a coherent framework for \textit{quantum wavelet shrinkage}, where sparse signal estimation is achieved through unitary wavelet transforms followed by channel-based or ancilla-driven attenuation. We construct forward and inverse quantum wavelet transforms as unitary operations, review circuit-level realizations inspired by the classical fast wavelet transform, and examine Givens-based decompositions \citep{givens1958} suitable for near-term hardware. We then introduce three complementary mechanisms for quantum shrinkage: (i) ancilla-driven CPTP channels that emulate thresholding and smooth shrinkage; (ii) damping-based channels that implement shrinkage through controlled decoherence; and (iii) hybrid quantum–classical schemes that reproduce blockwise or adaptive behavior through shallow entangling layers and post-measurement processing. 

By integrating these components, the framework connects classical adaptivity with quantum coherence. Sparsity and coherence, rather than conflicting, emerge as complementary aspects of information representation. The subsequent sections formalize these ideas, beginning with the unitary construction of quantum wavelet transforms and progressing to channel-based shrinkage models and their implementations on noisy intermediate-scale quantum (NISQ) hardware.

\section{Quantum Forward and Inverse Wavelet Transforms}
\label{sec:qwt}

Wavelet transforms in the quantum domain can be realized through several complementary constructions that strike a balance between algebraic rigor, circuit depth, and hardware feasibility. A straightforward option compiles the entire orthogonal wavelet transform into a single unitary that reproduces the full matrix exactly. An alternative approach emphasizes modularity, decomposing the operation into shallow layers of permutations and local rotations that mimic the classical fast wavelet transform (FWT). This includes expressing the transform as a structured sequence of Givens rotations directly matching the native gate sets of quantum processors. All approaches preserve orthogonality while trading off synthesis cost, circuit depth, and adaptability to different wavelet families.

Let $W \in \mathbb{R}^{N \times N}$ denote an orthogonal wavelet transform acting on a signal of length $N$.  
In the quantum setting, a real vector $x \in \mathbb{R}^N$ is encoded as
$$
|x\rangle = \frac{1}{\|x\|}\sum_{j=0}^{N-1} x_j\,|j\rangle,
$$
and the quantum wavelet transform (QWT) is a unitary $U_W$ satisfying
\begin{eqnarray}
U_W\,|x\rangle &=& |W x\rangle, \label{eq:QWT_forward}\\
U_W^\dagger\,|y\rangle &=& |W^\top y\rangle. \label{eq:QWT_inverse}
\end{eqnarray}
This mapping executes the entire multiscale decomposition coherently across superposition. For Daubechies, Coiflet, or Symlet families, it is defined by compactly supported quadrature mirror filters \citep[pp.~115-116]{vidakovic1999}. Classically, the FWT computes $W x$ through cascades of convolution and decimation in $O(N)$ time; the quantum task is to realize the same operation as an efficient unitary transformation acting on amplitudes.


\subsection{Direct Implementation as a Unitary Gate}
\label{subsec:direct_unitary}

The most direct realization of a QWT treats the full orthogonal wavelet matrix $W$ itself as a unitary operator. Because every real orthogonal matrix satisfies $W^\top W=I$, it defines a legitimate quantum gate $U_W$ acting as $U_W|e_j\rangle=|W e_j\rangle$. The transformation is executed coherently in a single step, with the inverse obtained immediately from $U_W^\dagger=U_{W^\top}$. This formulation is algebraically exact, perfectly reversible, and independent of auxiliary control or permutations, providing a reference for verifying more modular or approximate circuits.

The limitation is the synthesis cost. Decomposing a dense $N\times N$ matrix into native one- and two-qubit gates grows rapidly with $N$ and with filter length, particularly for long-support Daubechies or Symlet filters. Each choice of wavelet requires recompilation because the entries of $W$ depend explicitly on its coefficients. Nevertheless, the compiled unitary serves as a mathematically pristine baseline. On simulators and future fault-tolerant devices, it offers the most faithful realization of a QWT, single unitary gate embodying the entire orthogonal transformation.

\begin{example}  
This first example, featured in the Jupyter notebook {\tt QWShrink01.ipynb}, demonstrates that an orthogonal discrete wavelet transform can be expressed exactly as a single unitary quantum gate acting on amplitude-encoded data.
The notebook constructs the Daubechies DAUB2 wavelet transform (equivalent to the Haar transform but parameterized through the canonical Daubechies filter coefficients) as an $8\times8$ matrix applied to a three-qubit state vector.

The input sequence, as in \citet[p.113, Example 4.3.1]{vidakovic1999}, is transformed by this unitary operator through two decomposition levels, producing approximation and detail coefficients identical to those obtained by Mallat’s classical pyramid algorithm.
The notebook confirms numerical equality between the classical and quantum results, verifying that the $U_{\text{Daub2}}$ gate preserves orthogonality and energy exactly.

Conceptually, this example establishes the foundational link between classical wavelet matrices and quantum unitaries: every orthogonal wavelet transform can be implemented as a reversible quantum operation.
Subsequent examples in {\tt QWShrink02–06.ipynb} build upon this by embedding shrinkage and channel dynamics into the same framework.

\begin{figure}[h!]
\centering
\includegraphics[width=0.8\textwidth]{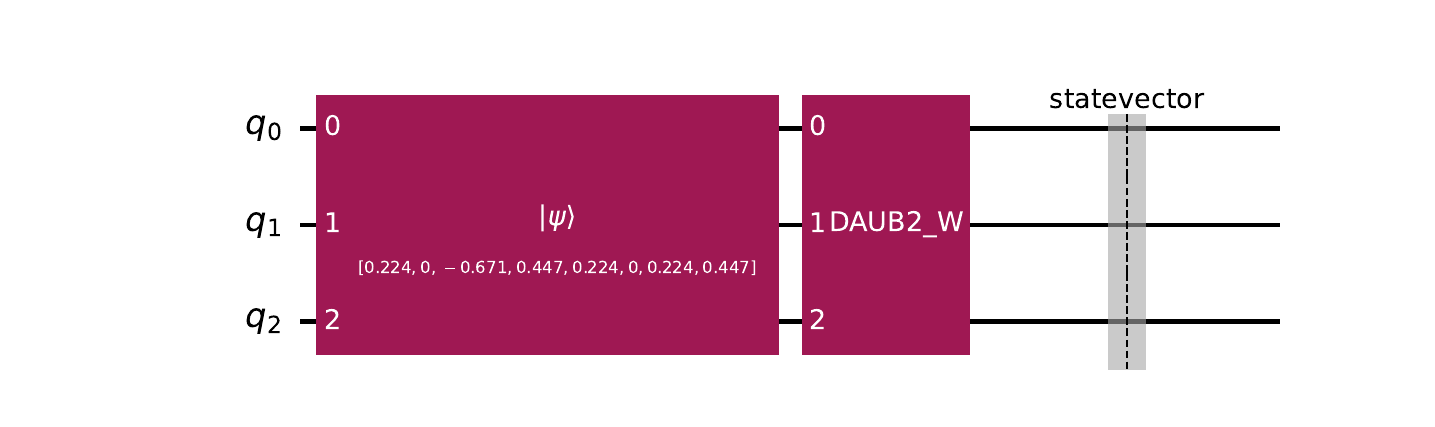} %
\caption{Quantum implementation of the Daubechies DAUB2 wavelet transform as a single $8\times 8$ unitary operation. 
The input sequence is amplitude-encoded into three qubits and transformed through two levels of decomposition. 
The resulting approximation and detail coefficients coincide exactly with those obtained by the classical wavelet-matrix and Mallat algorithms as in the example on pages~116-117 of \citet{vidakovic1999}, confirming the correctness and unitarity of the quantum realization.}
\label{fig:daub2_quantum}
\end{figure}

\end{example}

\subsection{Modular Quantum Realization via Filterbank Structure}
\label{subsec:compositional_qwt}

A more hardware-efficient construction mirrors the classical fast wavelet transform and the organization of the quantum Fourier transform. Here $U_W$ is composed of alternating permutation and rotation layers that act locally on subsets of qubits. Each permutation, such as a perfect shuffle or bit-reversal, is itself unitary and implemented efficiently by swap networks \citep{fijany1998springer}. These layers rearrange basis states between scales so that adjacent coefficients are paired for filtering.

Local rotations then perform the filtering. In the Haar case, the averaging and differencing steps correspond to $2\times2$ rotations acting on the subspaces $\{|2k\rangle,|2k+1\rangle\}$. For more general wavelets, the compactly supported filters give rise to $2\times2$ or $4\times4$ unitaries whose entries are determined by low-pass and high-pass coefficients. Such unitaries can be further decomposed into products of Givens rotations \citep{givens1958,hoyer1997,  fijany1998}. By combining these rotations with lightweight arithmetic for lifting-scheme predict–update steps, one obtains a modular ladder of blocks forming the complete $U_W$. Because many local gates act on disjoint qubit pairs, substantial parallelism is achievable, and the circuit depth scales polylogarithmically with $N$. This design thus provides an analytically transparent and hardware-aligned path to scalable quantum wavelet transforms.

\subsection{Givens-Based Quantum Wavelet Transforms}
\label{subsec:givens_based}

An even more structured representation is obtained by expressing $W$ as a product of elementary plane rotations, or Givens rotations \citep{givens1958}. Any orthogonal matrix $W\in\mathbb{R}^{N\times N}$ can be written as
$$
W = \prod_{k=1}^{K} G(i_k,j_k,\theta_k),
$$
where each $G(i,j,\theta)$ acts as identity except on $\text{span}\{e_i,e_j\}$, performing
$$
G(i,j,\theta)_{[i,j]} =
\begin{pmatrix}
\cos\theta & -\sin\theta\\
\sin\theta & \cos\theta
\end{pmatrix}.
$$
Each such two-level rotation is a native unitary operation on today’s quantum hardware, allowing direct physical realization. In this factorization, the Haar transform is a single rotation with $\theta=\pi/4$, and multi-level Haar transforms apply it in parallel to all disjoint pairs, followed by simple permutations that encode scale separation. More complex families (Daubechies, Symlets, Coiflets) yield small orthogonal blocks—typically $4\times4$ or $6\times6$, each decomposable into a few Givens rotations determined by the corresponding filter coefficients. The resulting QWT is a coherent cascade of local rotations interleaved with permutation gates, forming a reversible multiresolution hierarchy within the register. Because filter supports are short, the number of required rotations grows linearly with filter length and only logarithmically with signal size, maintaining low circuit depth while exactly preserving orthogonality. The Givens approach to quantum wavelet transform is featured in {\tt QWShrink03.ipynb} notebook.

\subsection{Complexity and Practical Considerations}
\label{subsec:cost}

State preparation remains a dominant cost in quantum signal processing, since encoding $x$ into amplitudes is typically $O(N)$. Once prepared, however, both modular and Givens-based circuits achieve polylogarithmic depth for structured transforms, similar to the quantum Fourier transform. Inversion is immediate because $U_W^\dagger$ is simply the reversed circuit with conjugated rotation angles. This reversibility is crucial for denoising applications, where the inverse transform reconstructs a state $|\tilde{x}\rangle$ after shrinkage in the wavelet domain.

Haar transforms admit extremely compact implementations using pairwise rotations and swaps. Longer-support wavelets require additional routing but remain feasible with structured decompositions \citep{LiLi2018,li2019,Nakahira2021,Bagherimehrab2023}. Linear-combination-of-unitaries methods provide a general mechanism for embedding arbitrary orthogonal matrices, though they usually entail a higher synthesis cost. Among the approaches discussed, the direct-unitary construction offers algebraic precision, the modular design ensures a scalable structure, and the Givens-based formulation aligns best with current hardware capabilities. Together, they span a flexible spectrum of strategies for implementing forward and inverse quantum wavelet transforms on near-term and fault-tolerant platforms.

\section{Quantum Realizations of Wavelet Shrinkage}
\label{sec:quantum_shrinkage_framework}

Classical wavelet shrinkage relies on nonlinear attenuation of noisy coefficients, a process that is straightforward in the classical domain but nontrivial to implement within the linear and unitary framework of quantum mechanics. To translate shrinkage into the quantum setting, one must reconcile three ingredients: the preservation of coherence, the physical constraints of quantum operations, and the adaptive suppression of noise-dominated coefficients. These considerations lead naturally to three distinct paradigms for quantum shrinkage, each grounded in a different physical mechanism and each suited to different hardware regimes.

The first paradigm uses ancilla-driven completely positive and trace preserving channels. Here, shrinkage is realized by embedding a nonunitary map inside a larger coherent unitary acting on an extended system, with attenuation emerging only when the ancilla is traced out. This approach is mathematically exact and fully reversible prior to taking the trace, and it allows fine-grained control over the amount of suppression applied to each coefficient.

The second paradigm exploits controlled decoherence, in which hardware-native phase damping serves as a direct physical surrogate for shrinkage. Coherence in selected directions is attenuated by a known factor determined by the device’s noise parameters. This strategy is simple, economical in qubits, and aligned with near-term hardware, though it is inherently irreversible once the phase information is lost.

The third paradigm consists of hybrid and alternative schemes. These include weak measurement, shallow entangling layers, feedback-controlled attenuation, and probabilistic uncomputation. Such methods preserve partial coherence while still enabling effective suppression of small coefficients. They require minimal additional resources and are particularly suitable for NISQ-era processors where circuit depth and coherence time are limited.

Together, these three approaches form a coherent framework for quantum wavelet shrinkage. They illustrate how the nonlinear idea of classical thresholding can be reformulated within quantum mechanics, with attenuation arising from coherent dilation, controlled decoherence, or adaptive hybrid dynamics depending on the physical capabilities of the device.


\subsection{Ancilla-Driven CPTP Shrinkage}
\label{subsec:ancilla_cptp}

Classical wavelet denoising modifies noisy coefficients through nonlinear shrinkage rules 
$S_\lambda(\cdot)$ such as  thresholding or smooth shrinkage policies 
\citep{donoho1994,donoho1995,vidakovic1999,VidakovicRuggeri2001}.
These rules suppress small coefficients and preserve large ones, but they do so through an irreversible and nonlinear transformation. A single quantum unitary cannot produce such an effect, since unitaries preserve norms and do not dissipate amplitude. If shrinkage is to be carried out in a quantum setting, it must arise as a physically valid quantum operation, and this leads naturally to completely positive and trace preserving maps implemented through Stinespring dilations.

\paragraph{CPTP maps and Kraus operators.}

A CPTP map $\mathcal{E}$ acting on a system $S$ may always be expressed in Stinespring form \citep{stinespring1955}. One introduces an ancilla or environment $E$, prepares it in a fixed pure state, applies a joint unitary evolution, and discards the ancilla. This beautifully simple picture captures how irreversible behavior arises from reversible quantum mechanics. The Stinespring representation is

\begin{eqnarray}
\mathcal{E}(\rho_S)
=
\operatorname{Tr}_E
\Big[
U_{SE}\,(\rho_S \otimes |0\rangle\langle 0|_E)\,U_{SE}^\dagger
\Big],
\label{eq:stinespring}
\end{eqnarray}
where $U_{SE}$ is a unitary on $S\otimes E$ and the partial trace removes the ancilla. The irreversible behavior of $\mathcal{E}$ comes entirely from ignoring the ancilla after the interaction.

To make this expression more concrete, suppose $\{|e_j\rangle_E\}_{j=0}^{d_E-1}$ is an orthonormal basis for the ancilla. The Kraus operators associated with $\mathcal{E}$ are

\begin{eqnarray}
K_j = \langle e_j |\, U_{SE}\, |0\rangle_E,
\qquad j=0,\dots,d_E-1,
\label{eq:kraus}
\end{eqnarray}
which satisfy the completeness condition

\begin{eqnarray}
\sum_{j=0}^{d_E-1} K_j^\dagger K_j = I.
\label{eq:kraus-complete}
\end{eqnarray}
The map itself becomes

\begin{eqnarray}
\mathcal{E}(\rho_S)
=
\sum_{j=0}^{d_E-1} K_j \rho_S K_j^\dagger.
\label{eq:kraus-sum}
\end{eqnarray}
Each Kraus operator represents how the system evolves conditioned on the ancilla ending in a particular state $|e_j\rangle_E$. In the context of shrinkage, we craft a pair of Kraus operators that contract the transverse Bloch components while leaving the longitudinal component untouched, which is exactly how classical smooth shrinkage behaves on the hidden states represented by wavelet coefficients.

\paragraph{Ancilla-driven implementation for wavelet coefficients.}

In a quantum wavelet transform, each coefficient $d_j$ is encoded as a small state $\rho_j$ on the coefficient register. To introduce shrinkage in a gentle and quantum-consistent way, we let $\rho_j$ interact with a fresh ancilla prepared in $|0\rangle_E$. Only the basis state $|j\rangle_S$ triggers a rotation on the ancilla. This very local and targeted interaction creates a small amount of entanglement. When we discard the ancilla, the remaining effect on $\rho_j$ becomes a CPTP map that acts like shrinkage.

The controlled transformation is

$$
|j\rangle_S|0\rangle_E
\longmapsto
|j\rangle_S
\big(
\sqrt{s_j}\,|0\rangle_E
+
\sqrt{1-s_j}\,|1\rangle_E
\big),
\qquad 0 \le s_j \le 1,
$$
while all other basis states of the coefficient register leave the ancilla unchanged. The parameter $s_j$ becomes the amount of shrinkage imposed on the $j$th coefficient.

After this interaction, we discard the ancilla. The resulting action on the coefficient register is

\begin{eqnarray}
\mathcal{E}_j(\rho_j)
=
s_j\,\rho_j
+
(1-s_j)\, Z\,\rho_j\,Z,
\label{eq:ancilla-channel}
\end{eqnarray}
where $Z$ is the Pauli $Z$ operator. A convenient Kraus representation is

$$
K_0 = \sqrt{s_j}\, I, \qquad 
K_1 = \sqrt{1-s_j}\, Z,
$$
with $K_0^\dagger K_0 + K_1^\dagger K_1 = I$ as required. In geometric terms, the longitudinal Bloch component remains unchanged while the transverse components are contracted. This contraction is exactly the type of continuous shrinkage that classical soft thresholding performs. Smaller values of $s_j$ correspond to stronger attenuation.

\paragraph{Unitary representation and control angles.}

The joint unitary that gives rise to the channel (\ref{eq:ancilla-channel}) satisfies a compact expression. Let $X_E$ denote the Pauli $X$ acting on the ancilla. Then

\begin{eqnarray}
U_{SE}
=
\exp
\Big[
 -i
 \sum_j
 \theta_j
 \big(
   |j\rangle\langle j|_S \otimes X_E
 \big)
\Big],
\label{eq:ancilla-unitary}
\end{eqnarray}
where the rotation angle $\theta_j$ determines the shrinkage strength. The relationship

\begin{eqnarray}
s_j = \cos^2(\theta_j)
\label{eq:sj-theta}
\end{eqnarray}
connects the two. By choosing $\theta_j$ to depend on the wavelet scale or position, one recovers traditional levelwise or blockwise shrinkage strategies. The important point is that all evolution is coherent until the ancilla is traced out. The nonlinearity enters only at that last step, just as it should.

This ancilla-driven scheme requires one ancilla per coefficient, or one per block of coefficients if the same shrinkage level is applied across a block. The method also needs controlled rotations whose angles depend on the coefficient index. Although this introduces overhead, it yields a principled and physically compatible quantum mechanism for shrinkage. All operations remain linear and unitary on an enlarged space, and the nonlinearity appears only upon tracing out the ancilla, which is exactly how CPTP maps arise in quantum mechanics.

\subsection{Controlled Decoherence as Shrinkage Surrogate}
\label{subsec:phase_damping}

A more hardware-aligned alternative uses intrinsic noise channels as computational primitives.  
The phase-damping (pure dephasing) channel acts on a single-qubit density matrix $\rho$ as
$$
\mathcal{E}_\gamma(\rho)
= K_0\rho K_0^\dagger + K_1\rho K_1^\dagger,
\qquad
K_0=
\begin{bmatrix}
1&0\\0&\sqrt{1-\gamma}
\end{bmatrix},
\quad
K_1=
\begin{bmatrix}
0&0\\0&\sqrt{\gamma}
\end{bmatrix},
$$
where $\gamma\in[0,1]$ is the dephasing strength \citep{nielsen2010,lidar2013}.  
In Bloch-vector form $(x,y,z)$, this yields
$$
(x,y,z)\;\mapsto\;(\sqrt{1-\gamma}\,x,\;\sqrt{1-\gamma}\,y,\;z),
$$
so that coherence in the $xy$-plane is reduced by $\sqrt{1-\gamma}$ while populations remain unchanged.  
If a coefficient $d_i$ is encoded as $\langle X_i\rangle=x_i$, then
\begin{eqnarray}
\langle X_i\rangle_{\text{after}}
= \sqrt{1-\gamma_i}\,\langle X_i\rangle_{\text{before}}
= \sqrt{1-\gamma_i}\,x_i.
\label{eq:sqrt1mgamma}
\end{eqnarray}
Thus, the transformation
$$
x_i\;\longmapsto\;\sqrt{1-\gamma_i}\,x_i
$$
acts as multiplicative shrinkage.  
Choosing $\gamma_i$ in (\ref{eq:sqrt1mgamma}) as a function of $|x_i|$ emulates classical hard thresholding, $\gamma_i ={\mathbf 1}(|x_i|\leq \lambda),$ 
or for a smooth rule, $\gamma_i=1-\exp(-\alpha|x_i|)$ or $\gamma_i = \cos\left(\frac{\pi}{2} |x|^\alpha\right)$ (as in Fig.\ref{fig:phase_soft} for $\alpha=4$).



This mechanism is physically transparent and very easy to implement on current hardware, since dephasing is a native noise process and already acts as a legitimate CPTP channel on the data qubit. No ancillary systems are required to generate the shrinkage itself, because the attenuation factor $\sqrt{1-\gamma}$ arises directly from genuine decoherence rather than from tracing out a larger coherent system. If an ancilla appears in circuit diagrams for this method, it functions only as a classical control line that selects when the dephasing channel is applied, not as a subsystem whose trace produces the CPTP map. The approach integrates naturally with a quantum wavelet transform circuit and is attractive for near-term devices. At the same time, it is fundamentally irreversible: once phase information is lost, it cannot be recovered. Thus, although controlled dephasing provides a simple and hardware-aligned surrogate for classical shrinkage, it lacks the full reversibility, precision, and tunability of the ancilla-driven model.

\vspace*{0.2in}

\begin{example}
This example illustrates how a simple phase damping channel can reproduce both a hard type thresholding rule and a smooth shrinkage rule, using only a small register of three qubits. No ancilla qubits are used. The starting point is the coefficient vector
$ 
d = (2, 1, 9, 0, 3, -10, 2, 4),
$
which is first rescaled affinely to the interval $[-1,1]$. The rescaled values $d_i$ serve as target expectations $\langle X\rangle$ for single qubits. Each value $d_i$ is encoded into a three qubit product state as follows. All qubits are prepared in $|+\rangle^{\otimes 3}$, and then a rotation $R_z(\phi_i)$ with $\cos(\phi_i)=d_i$ is applied to every qubit. After this step the ideal expectation of the Pauli $X$ operator on each qubit equals $d_i$, and the average over the three qubits recovers $d_i$ up to sampling noise.

For the hard type threshold we fix $\lambda=0.4$ on the rescaled scale and define the coefficientwise phase damping parameter
$\gamma_i = \mathbf{1}(|d_i| \leq \lambda).$
Large coefficients experience no decoherence, while small coefficients are fully dephased. The phase damping channel with parameter $\gamma_i$ is applied independently to each qubit. The circuit is closed with another layer of Hadamard gates and measurement in the computational basis. The empirical average of $\langle X\rangle$ over the three qubits and many shots yields the observed shrunk coefficient. Ideally, the channel transforms $\langle X\rangle$ as
$$
\langle X\rangle \mapsto \sqrt{1-\gamma_i}\,\langle X\rangle ,
$$
so in this binary case the theoretical output is either $d_i$ (when $|d_i|>\lambda$) or $0$ (when $|d_i|\le\lambda$). The first accompanying figure shows the original rescaled coefficients, the simulated shrunk values, and the ideal hard threshold profile that keeps only those entries with $|d_i|>\lambda$, as in Fig. \ref{fig:phase_threshold}.

In the second panel the discontinuous threshold is replaced by a smooth shrinkage rule, again without ancilla. The phase damping parameter is chosen as
$$
\gamma_i = \cos^{4}\!\left( \frac{\pi}{2}\,|d_i| \right),
$$
so that $\gamma_i$ is close to $1$ near the origin and approaches $0$ as $|d_i|$ approaches $1$. Under this channel the ideal transformation of the encoded expectation is
$$
d_i \mapsto \sqrt{1-\gamma_i}\, d_i ,
$$
which continuously attenuates smaller coefficients while leaving the large ones approximately unchanged, as in Fig.\ref{fig:phase_soft} 

The circuit is the same as before, and the measured curve follows the theoretical profile with small perturbations due to shot noise. The second figure shows the original coefficients, the ideal curve $\sqrt{1-\gamma_i}\,d_i$, and a simulated measurement trace.

These two constructions, featured in {\tt QWShrink07.ipynb}, demonstrate that by encoding coefficients into $\langle X\rangle$ and selecting suitable phase damping parameters $\gamma_i$ one can realize both an abrupt hard threshold rule and a smooth shrinkage rule as completely positive trace preserving maps on a small register. The implementation is ancilla free and uses only local Kraus channels acting on the data qubits.

\begin{figure}[h!]
\centering
\includegraphics[width=0.8\textwidth]{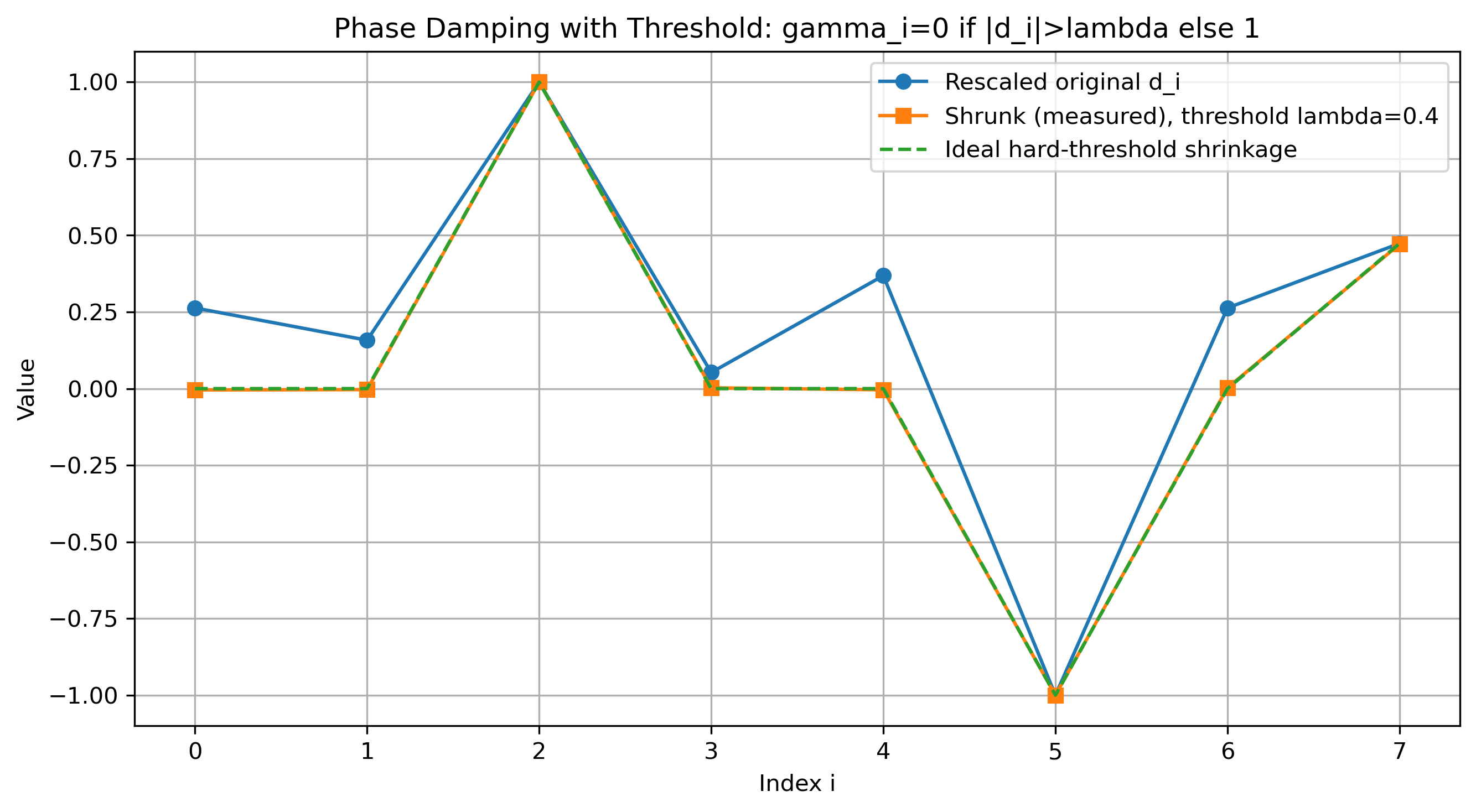} %
\caption{\small Phase-based thresholding map in the quantum wavelet framework. We start with wavelet coefficients
$d=[2, 1, 9, 0, 3, -10, 2, 4]$ and rescale them to 
$[-1,1].$  This short sequence of classical values is used throughout this paper 
to zoom on the local action of wavelet shrinkage. The compact example makes each step of the transformation easy to follow, while the same procedures apply without difficulty to sequences of length in the thousands.
Each rescaled wavelet coefficient undergoes a phase rotation determined by its magnitude, producing an effective thresholding rule in which small coefficients are strongly displaced toward destructive interference while large coefficients remain nearly unchanged.
The resulting mapping provides a fully unitary realization of coefficient shrinkage.}
\label{fig:phase_threshold}
\end{figure}
 
\begin{figure}[h!]
\centering
\includegraphics[width=0.8\textwidth]{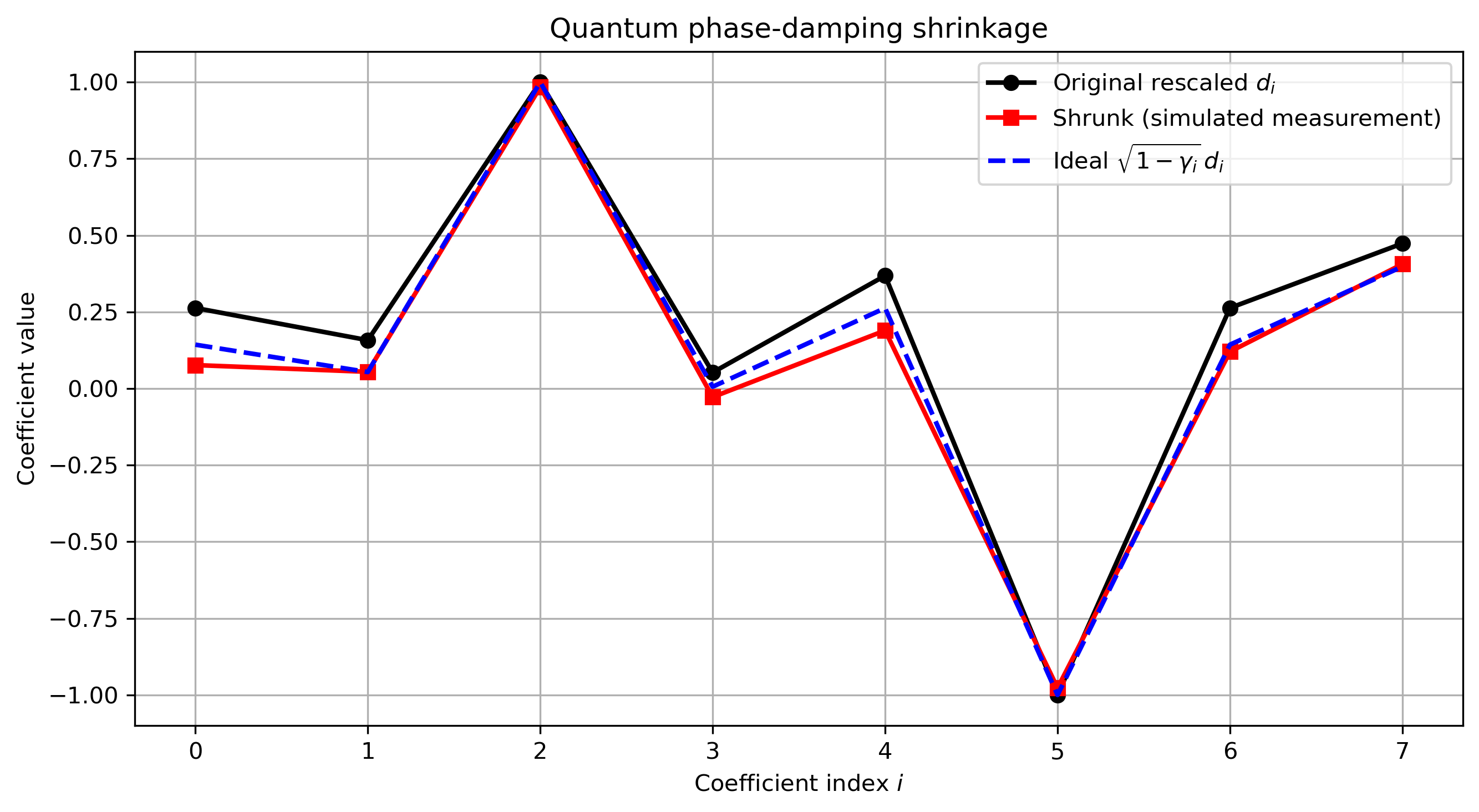} %
\caption{\small  Soft-shrinkage behavior emerging from phase-controlled attenuation.
The plot relates the input and output amplitudes under the phase-driven rule, showing a smooth contraction that parallels the classical soft-threshold function.
This unitary realization anticipates the probabilistic shrinkage of later Kraus and CPTP channel formulations as in Subsection \ref{subsec:phase_damping}.}
\label{fig:phase_soft}
\end{figure}

\end{example}

\subsection{Hybrid and Feedback-Assisted Shrinkage Mechanisms}
\label{subsec:hybrid_shrinkage}

Between the fully coherent ancilla-driven CPTP model and the hardware-native dephasing surrogate lies a wide middle ground of hybrid techniques. These strategies use weak measurement, partial feedback, or shallow entangling layers to achieve controlled attenuation while still preserving some degree of quantum coherence. They are flexible, intuitive, and particularly attractive for NISQ-era demonstrations, where one often wishes to temper the strength of decoherence without relying on large numbers of ancillary qubits or deep circuits.

\paragraph{Weak measurement.}

A weak measurement extracts only partial information about a coefficient observable $A_j$ and therefore avoids a full projective collapse. This permits a gentle, tunable form of shrinkage. A simple model is

$$
\rho_j
\longmapsto
(1-\eta)\rho_j + \eta\, M_j \rho_j M_j^\dagger,
$$
where $\eta\in[0,1]$ controls the measurement strength and $M_j$ is the corresponding measurement operator \citep{aharonov1988,jacobs2014}. When $\eta$ is small, the state remains largely coherent and only slightly nudged toward the measurement outcome; when $\eta$ approaches one, the operation approaches a full projection. By selecting $\eta$ as a function of the estimated noise level or coefficient magnitude, small coefficients experience stronger partial collapse and large coefficients remain nearly untouched. In this way, weak measurement provides a quantum analogue of soft thresholding: it interpolates smoothly between identity and strong suppression without fully sacrificing coherence.

\begin{example}
To illustrate the effect of quantum wavelet shrinkage, the accompanying notebook \texttt{QWShrink05.ipynb} generates the Donoho–Johnstone Doppler test-signal and applies controlled noise with SNR of $7$. The quantum denoising step is implemented as a CPTP channel that attenuates wavelet coefficients coherently, emulating soft thresholding in a quantum setting. The notebook visualizes three signals: the original (clean) Doppler, its noisy counterpart, and the reconstruction obtained by the CPTP channel. These results demonstrate how the method described in Section~\ref{subsec:ancilla_cptp} achieves effective denoising while maintaining the oscillatory structure of the Doppler signal.

\begin{figure}[h!]
\centering
\includegraphics[width=0.95\textwidth]{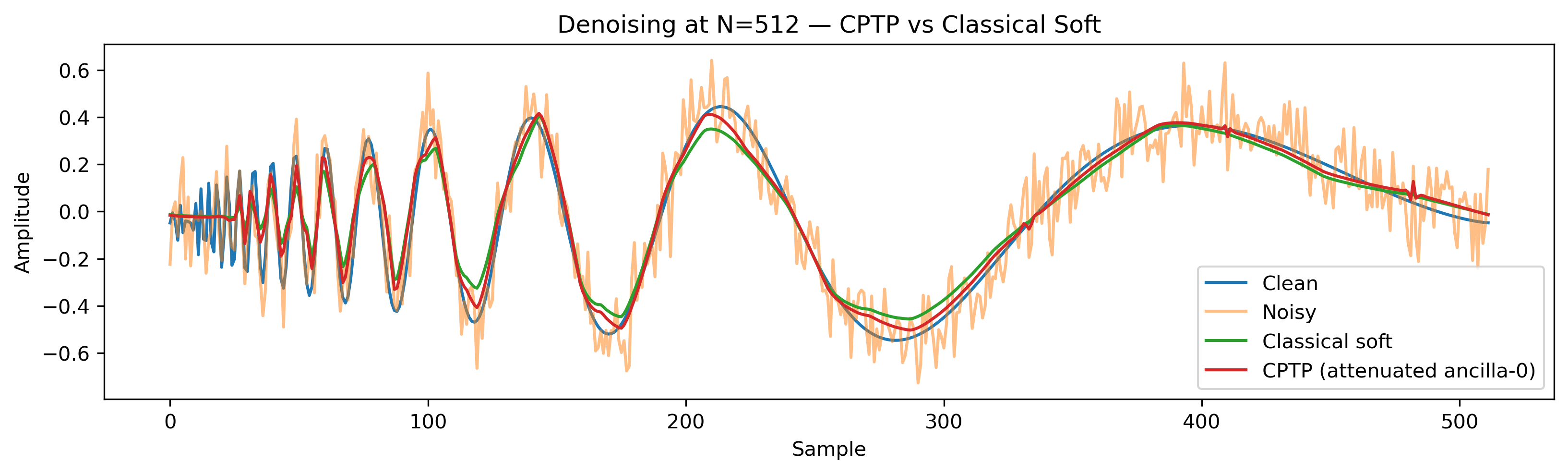} %
\caption{\small  Comparison of classical soft thresholding with CPTP-based shrinkage. The clean Doppler signal, its noisy version, the reconstruction via classical soft thresholding, and the reconstruction via CPTP attenuation of wavelet coefficients are shown. The CPTP method uses ancilla-0 diagonal scaling to impose coefficientwise attenuation that mimics soft shrinkage while remaining physically admissible as a quantum channel. The construction of
the CPTP shrinkage channel is described in Section~\ref{subsec:ancilla_cptp}.}
\label{fig:doppler_cptp}
\end{figure}

\noindent
The present CPTP formulation expresses shrinkage directly as a density–matrix attenuation channel, 
rather than as a sum of explicit operator actions. 
This abstraction allows parameterized control of the shrinkage strength and provides a convenient bridge between physical realization and statistical modeling.

\begin{figure}[h!]
\centering
\includegraphics[width=0.8\textwidth]{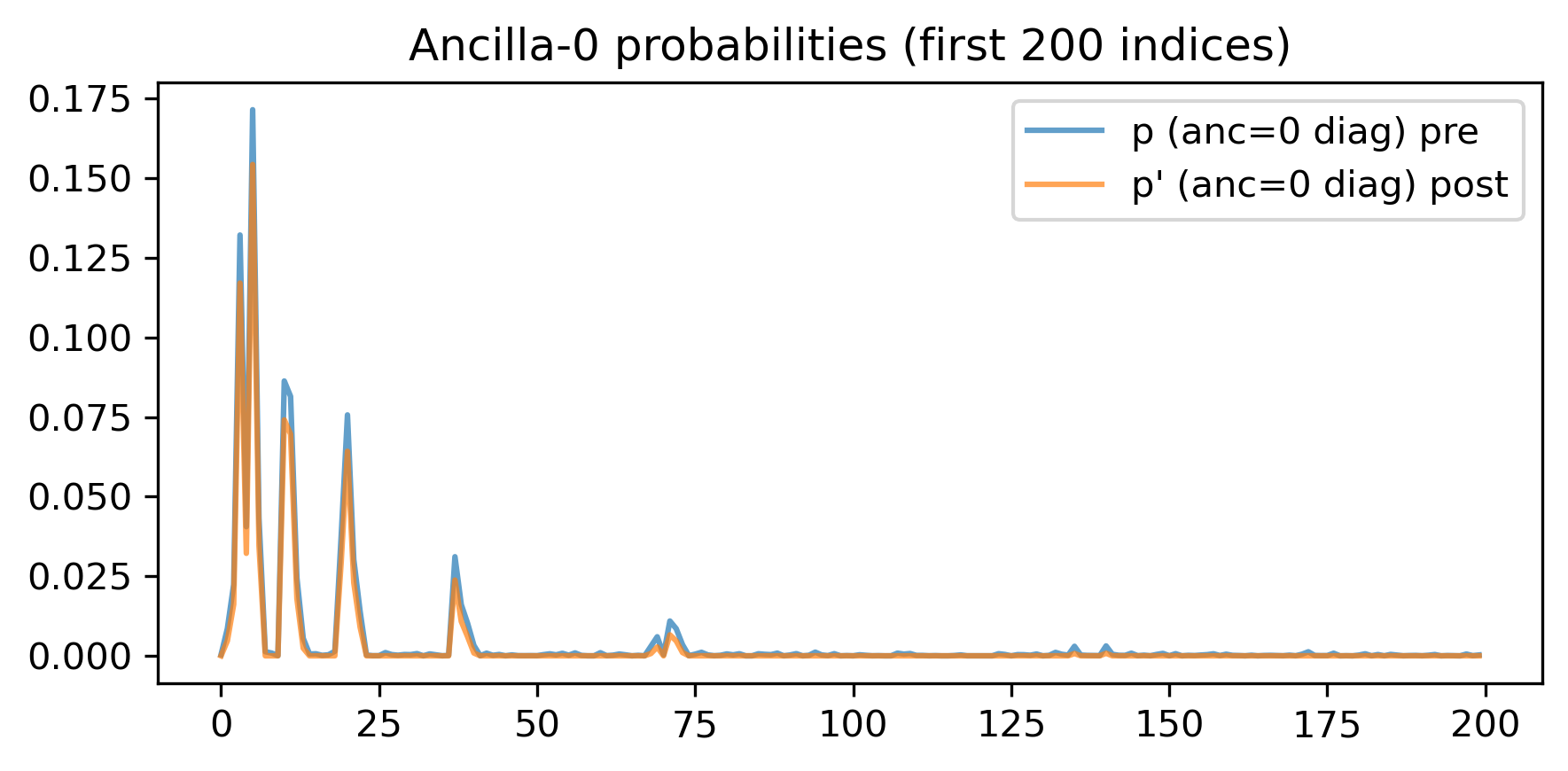} %
\caption{\small
Ancilla-0 diagonal probabilities before and after CPTP attenuation. The plot displays the first 200 indices of the diagonal of the reduced density matrix associated with ancilla outcome zero. The post-attenuation curve shows the multiplicative action of the CPTP channel, confirming that the channel scales each coefficient level according to the prescribed attenuation factors.
}
\label{fig:e05CTP3}
\end{figure}

\noindent
Figure \ref{fig:e05CTP3} illustrates how CPTP attenuation channel redistributes probability between the data register and the ancillary system. 
The blue curve (\textit{pre}) represents the diagonal elements of the reduced density matrix corresponding to the ancilla state $|0\rangle$ before the channel acts—essentially the undisturbed, pre-interaction population, which is nearly uniform because the ancilla is initialized in $|0\rangle$. 
The orange curve (\textit{post}) shows the same probabilities after the CPTP map has been applied. 
Their difference quantifies the local strength of shrinkage: where the orange curve dips below the blue, probability amplitude has been transferred from the signal register to the ancilla’s $|1\rangle$ state, indicating stronger damping of the corresponding wavelet coefficients. 
This visualization provides an internal diagnostic of how the quantum channel enacts shrinkage across the coefficient index space.
\end{example}

\paragraph{Neighborhood mixing.~}
Another hybrid approach uses shallow entangling layers to mix neighboring coefficients prior to any measurement or decoherence step. A simple mixing unitary is

$$
U_{\text{mix}}(\alpha)
=
\prod_j
\exp\big[ -i\alpha\, (X_j X_{j+1} + Y_j Y_{j+1}) \big],
$$
which produces effective coefficients of the form $\tilde d_j \approx d_j + \alpha(d_{j-1}+d_{j+1})$ when $\alpha$ is small. This creates local correlations and spreads information among adjacent wavelet coefficients. Classical shrinkage or thresholding applied to the mixed coefficients $\tilde d_j$ then mimics block or neighborhood-based shrinkage methods \citep{cai1999,vidakovic1999}. The important point is that coherence is preserved throughout the mixing stage; the only irreversible step appears at the final classical or measurement stage, giving this method a pleasant hybrid character.

\begin{example}
 This example introduces smooth probabilistic ancilla shrinkage in which each coefficient is attenuated according to a data-dependent probability $p_i$, producing the transformation $x_i \mapsto p_i x_i$. While the method does not explicitly mix neighboring coefficients, it represents the same intermediate regime: attenuation is nonunitary but remains smooth and adaptive, and coherence is only partially lost.
Here, the shrinkage law $f(x)$ can be viewed as a target amplitude-damping profile, suitable for later realization via CPTP or Kraus channels.
It represents a deterministic shrinkage design stage rather than a physical quantum process.

\begin{figure}[h!]
\centering
\begin{minipage}[t]{0.48\textwidth}
    \centering
    \includegraphics[width=\textwidth]{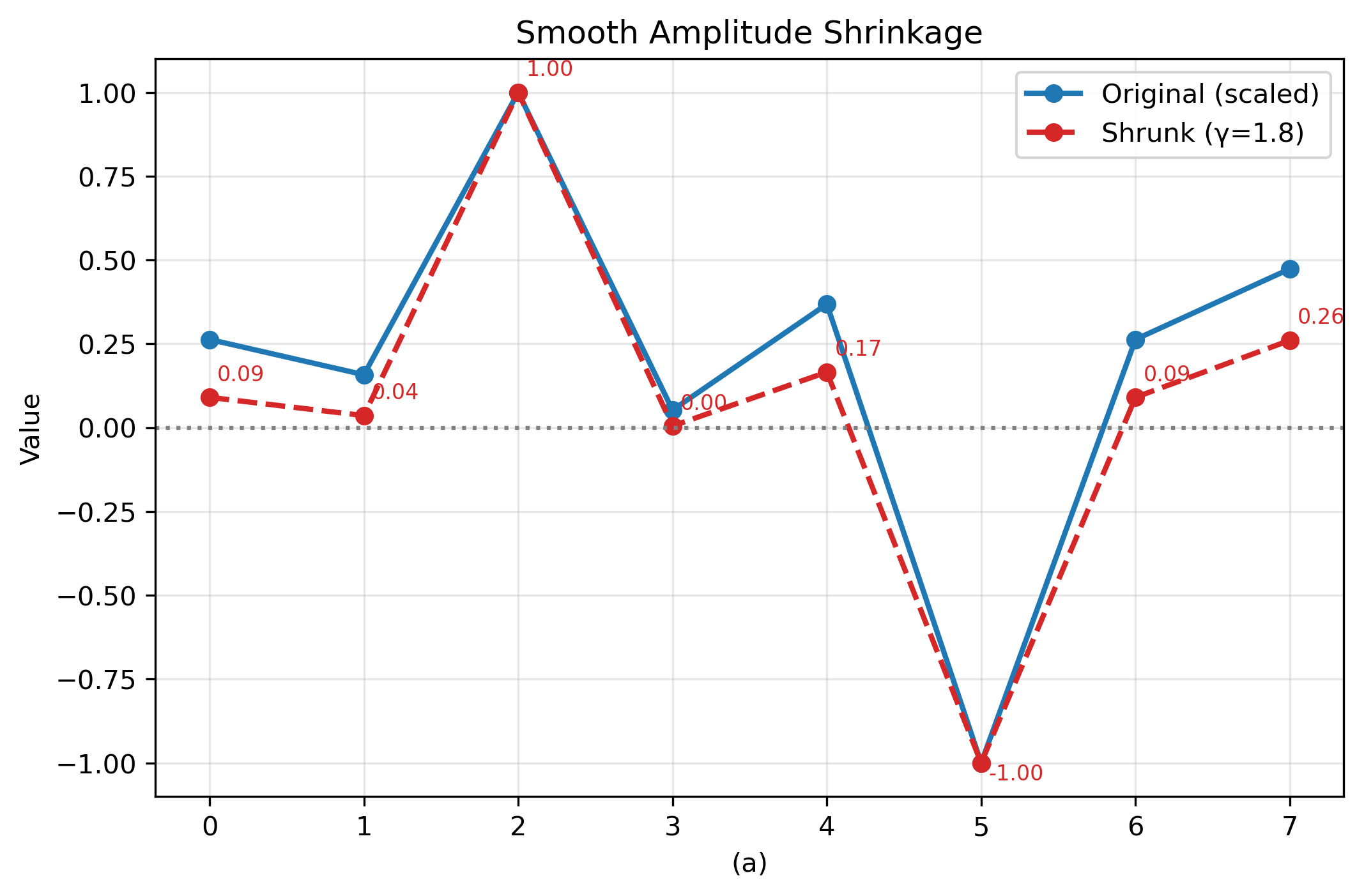}
\end{minipage}
\hfill
\begin{minipage}[t]{0.48\textwidth}
    \centering
    \includegraphics[width=\textwidth]{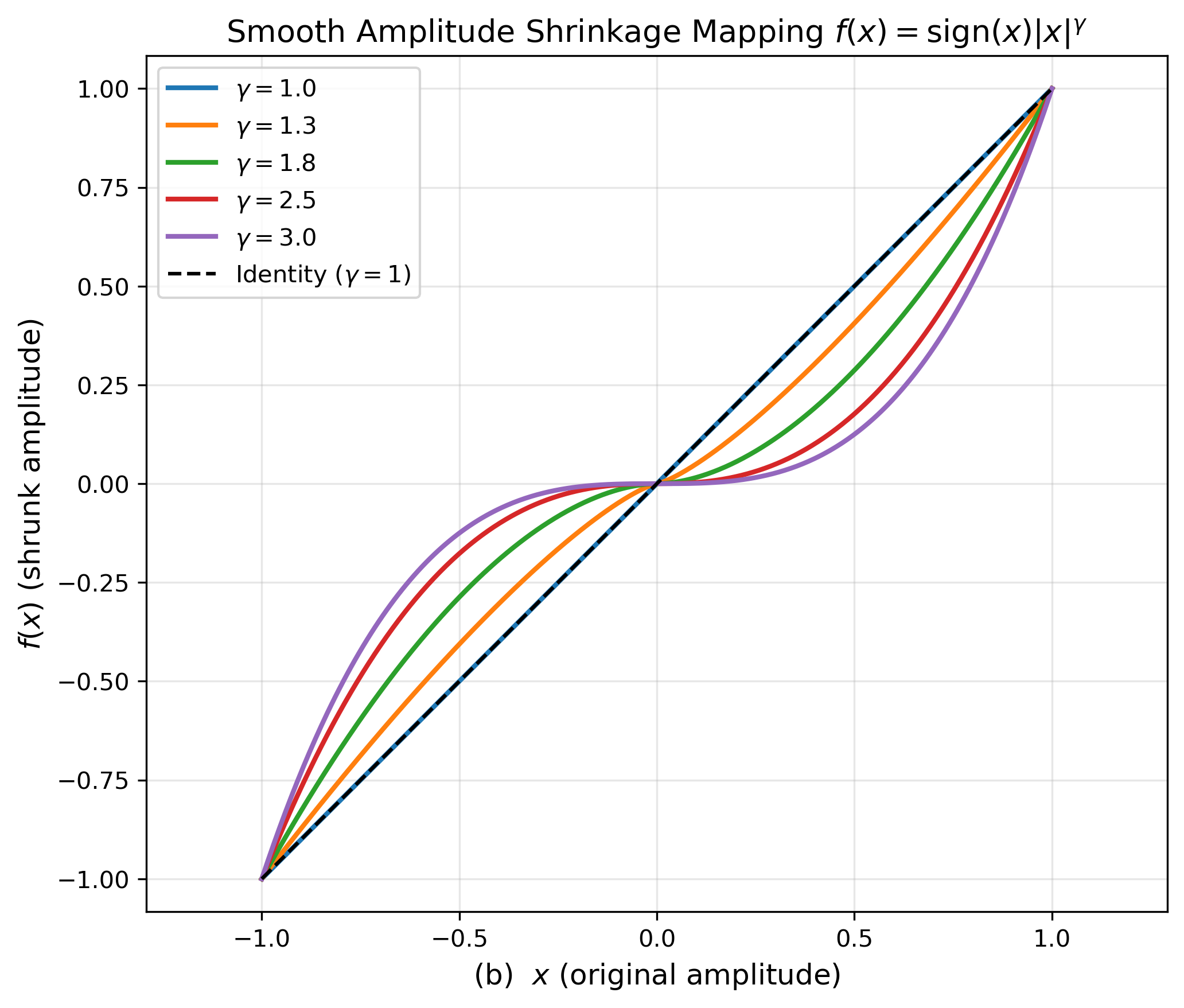}
\end{minipage}
\caption{\small Panel~(a) shows the application of the mapping 
$f(x)=\operatorname{sign}(x)|x|^{\gamma}$ with $\gamma=1.8$ 
to a representative signal, producing gradual amplitude damping that preserves large features while suppressing small ones.
Panel~(b) depicts the shrinkage rule itself, illustrating how the power parameter $\gamma$ controls the curvature and smoothness of the attenuation.}
\label{fig:ex10smoothrule}
\end{figure}
\end{example}

\paragraph{Feedback and postselection.}

A more dynamic mechanism combines weak measurement with conditional unitary rotations. A general feedback map takes the form

$$
\rho_j
\longmapsto
\sum_m R_m\, M_m \rho_j M_m^\dagger R_m^\dagger,
$$
where $M_m$ are weak measurement operators and $R_m$ are conditional rotations applied after each outcome \citep{wiseman2009}. This scheme can attenuate small coefficients more aggressively by applying stronger corrective rotations in outcomes associated with weaker signal components. Alternatively, one may employ postselection-based protocols: after a weak interaction, an ancilla is reset or uncomputed only when a desired outcome is obtained. Although probabilistic, such non-deterministic schemes approximate nonlinear shrinkage without requiring full projective measurement or heavy decoherence. They sit between coherent control and irreversible suppression, and allow a degree of adaptivity that can be quite useful in small-scale demonstrations.

\begin{example} This example featured in {\tt QWShrink08.ipynb} demonstrates a minimal ancilla-driven implementation of quantum shrinkage,
where a secondary qubit acts as a flag that identifies coefficients exceeding a selected
threshold. We are back to the short sequence of classical values $[2, 1, 9, 0, 3, -10, 2, 4]$.  The sequence is first
normalized to $[-1, 1]$ and each element is amplitude-encoded as a rotation on a
single-qubit register. The ancilla qubit is conditionally flipped whenever the magnitude
of the encoded coefficient exceeds a preassigned threshold $\lambda = 0.4$. Measurement
of the ancilla then yields a probability $P(\text{flag}=1)$ indicating whether the
coefficient should be retained or shrunk as in Fig. \ref{fig:e08AncilaryThresholding}(b)

Running the circuit for each coefficient produces a probabilistic analog of
classical hard thresholding: large coefficients correspond to high ancilla excitation
probabilities, while smaller coefficients are suppressed. The results are visualized in
two panels in Fig. \ref{fig:e08AncilaryThresholding}. The left panel displays the normalized data vector, and the right panel
shows the corresponding ancilla-activation probabilities estimated from $1024$ simulated
shots using the \texttt{Qiskit~2.x} \texttt{AerSimulator}. Together, they provide an
explicit example of ancilla-driven shrinkage, in which controlled decoherence replaces
deterministic thresholding.

\begin{figure}[h!]
\centering
\includegraphics[width=0.95\textwidth]{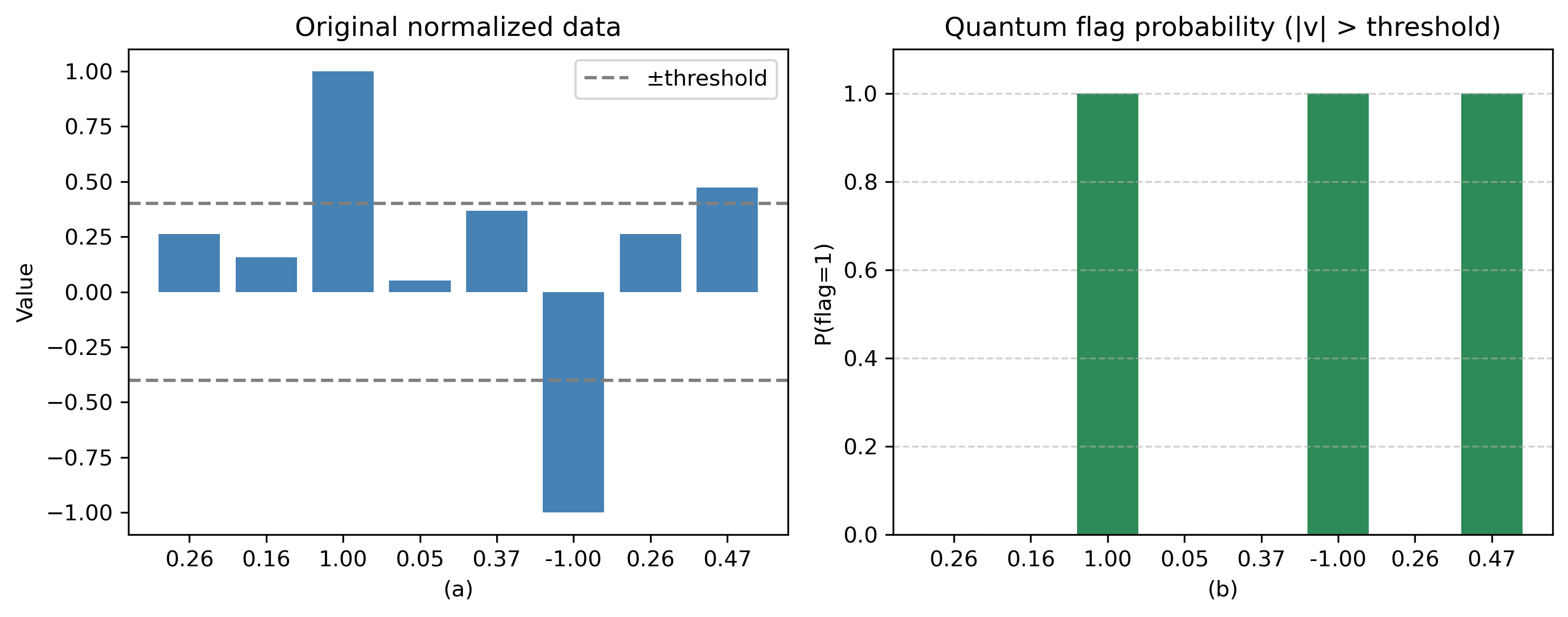} %
\caption{\small 
(a) Original normalized wavelet coefficients $d_j$ obtained from the sequence
$[2, 1, 9, 0, 3, -10, 2, 4]$. The dashed lines at $\pm \lambda = 0.4$ mark
the classical shrinkage thresholds;~
(b) Ancilla-driven shrinkage results. The bars indicate the measured probability
$P(\text{flag}=1)$ for each coefficient, representing the likelihood that the coefficient
exceeds the threshold. Larger coefficients activate the ancilla more strongly,
illustrating a probabilistic quantum surrogate for hard thresholding.}
\label{fig:e08AncilaryThresholding}
\end{figure}
\end{example}

\begin{example}
 In this example ({\tt QWShrink09.ipynb}), a sequence of normalized coefficients is processed through a
two-qubit circuit in which the ancilla qubit acts as a continuously controlled attenuator.
Unlike the binary activation used for hard thresholding, the ancilla rotation angle here
is proportional to the coefficient magnitude, producing a smooth range of excitation
probabilities between $0$ and $1$
(Fig. \ref{fig:e09ancsoft}(b)).
The resulting expectation value of the ancilla
represents a continuous shrinkage factor that weakens small coefficients and preserves
large ones, a quantum analog of the classical smooth shrinkage rule.

This experiment fits smoothly into the theoretical framework developed in this section. Completely positive trace preserving maps were introduced as coherent substitutes for nonlinear shrinkage rules, and the ancilla in this example offers a concrete realization of that viewpoint. Its controlled rotation prepares the desired balance of amplitudes, and the partial measurement transfers just enough probability to produce a Kraus-weighted attenuation on the main register. The level of shrinkage is carried directly by the excitation probability of the ancilla. In this way the smooth shrinkage effect arises naturally from the same physical principles that underlie quantum decoherence and partial entanglement.

\begin{figure}[h!]
\centering
\includegraphics[width=0.95\textwidth]{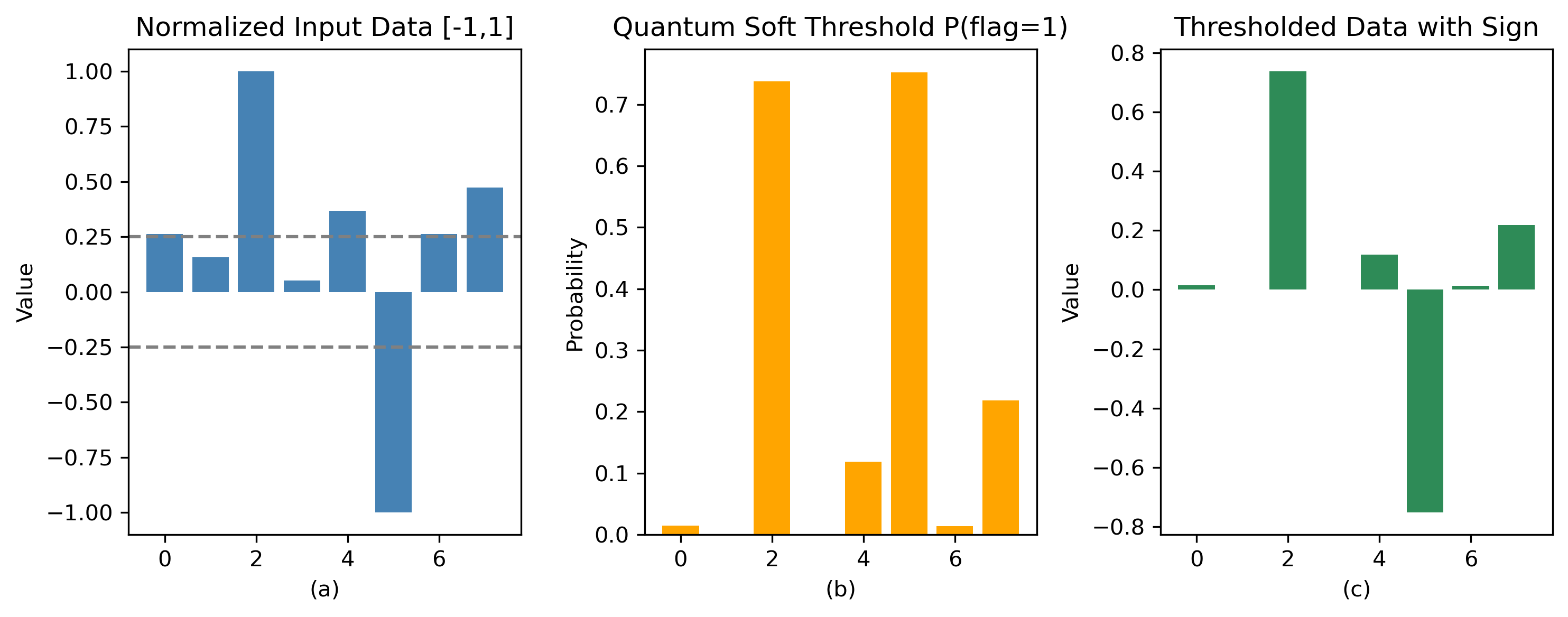} %
\caption{\small(a) Normalized input coefficients used for smooth thresholding demonstration.
The data are scaled to $[-1,1]$ prior to amplitude encoding, with dashed lines indicating
the nominal shrinkage threshold. (b) Ancilla expectation values corresponding to each coefficient.
The continuous variation of $\langle Z \rangle_{\text{ancilla}}$ between 0 and 1 represents
a smooth attenuation curve, analogous to classical smooth shrinkage. Larger coefficients
maintain near-unity ancilla excitation, while smaller ones are progressively suppressed. (c) Shrunk coefficients.}
\label{fig:e09ancsoft}
\end{figure}

\end{example}

\subsection{Comparative Discussion}
\label{subsec:comparison}

The three shrinkage mechanisms form a natural spectrum of quantum realizations for wavelet attenuation. At one end, the ancilla-driven CPTP construction is mathematically exact and fully coherent until the ancilla is traced out. It offers precise and tunable shrinkage through controlled rotations but requires additional qubits and accurate control over multiqubit interactions. At the opposite end, phase-damping attenuation relies on the hardware’s native dephasing channel. It is extremely simple to implement and integrates seamlessly with a quantum wavelet transform, yet it is fundamentally irreversible and offers less flexibility in shaping the attenuation profile. Hybrid strategies occupy the middle ground. They rely on weak measurement, shallow entangling layers, or feedback-conditioned rotations to introduce controlled suppression while preserving some degree of coherence. These schemes demand fewer resources than the ancilla-based model and remain more reversible and tunable than raw decoherence, making them attractive for NISQ-era experiments.

\medskip

Taken together, these approaches show how the essence of wavelet shrinkage, namely the adaptive suppression of noise-dominated coefficients, can be reinterpreted in quantum mechanical terms. What is nonlinear in the classical domain becomes linear and trace preserving once the system is viewed as part of a larger or open quantum environment, and adaptivity arises naturally through controlled decoherence, weak measurement, or feedback. In this sense, quantum wavelet shrinkage highlights a productive intersection of statistical inference and quantum dynamics: attenuation, coherence, and sparsity become aspects of the same physical principle when viewed through the lens of quantum operations.

\section{Encoding Classical Data for Quantum Wavelet Shrinkage}
\label{sec:encoding}

Quantum wavelet shrinkage relies on mappings from classical numerical quantities to quantum states. The form of this encoding governs which unitaries or channels are natural, how attenuation manifests as shrinkage, and how denoised estimates are recovered after measurement \citep{SchuldPetruccione2018,BiamonteEtAl2017}. In classical denoising, wavelet coefficients are explicit numerical scalars; in the quantum domain they become amplitudes, phases, or expectation values of qubit states. This section, which follows the development of quantum shrinkage mechanisms in Section~\ref{sec:quantum_shrinkage_framework}
clarifies how encoding interacts with the ancilla-driven, decoherence-driven, and hybrid methods introduced earlier, and how these encodings are implemented in the some of the examples.

\vspace*{0.15in}
\noindent {\bf Amplitude Encoding.}
Amplitude encoding maps a real or complex vector $x=(x_0,\ldots,x_{N-1})$ into a normalized quantum state
\begin{eqnarray}
|x\rangle=\frac{1}{\|x\|}\sum_{j=0}^{N-1}x_j|j\rangle,
\end{eqnarray}
so that each data component appears as a quantum amplitude \citep{SchuldPetruccione2018,BiamonteEtAl2017}. If $U_W$ is the unitary implementing an orthogonal wavelet transform $W$, then
\begin{eqnarray}
U_W|x\rangle = |W x\rangle,
\end{eqnarray}
and the wavelet coefficients are again stored as amplitudes.

This encoding aligns directly with the coherent mechanisms described in Sections~\ref{subsec:ancilla_cptp} and~\ref{subsec:hybrid_shrinkage}, since orthogonal matrices lift naturally to unitary operators. It is therefore well suited for ancilla-driven or Kraus-based CPTP shrinkage, where attenuation is achieved through coherent unitary interactions on an extended Hilbert space. Challenges include the global normalization constraint $\sum_j |x_j|^2 = 1$ and the fact that amplitude measurement is destructive. As a result, amplitude encoding is most powerful when shrinkage is performed coherently and the full quantum state is preserved until the final readout.

In the ancilla-driven CPTP setting, attenuation factors $s_j$ act multiplicatively on amplitudes:
\begin{eqnarray}
|j\rangle|0\rangle_a \mapsto |j\rangle \big( \sqrt{s_j}|0\rangle_a + \sqrt{1-s_j}|1\rangle_a \big),
\end{eqnarray}
so that discarding the ancilla produces the effective transformation $x_j \mapsto \sqrt{s_j}\,x_j$. This realizes wavelet shrinkage as a coherent dilation followed by a trace, in full agreement with the CPTP formulation of Section~\ref{subsec:ancilla_cptp}.

\vspace*{0.15in}
\noindent {\bf Phase or Expectation-Value Encoding.}
A different representation stores numerical coefficients in expectation values of observables such as Pauli operators. For a scalar $x_i \in [-1,1]$, one prepares a one-qubit state $|\psi_i\rangle$ satisfying
\begin{eqnarray}
\langle \psi_i|X|\psi_i\rangle = x_i,
\end{eqnarray}
where $X$ is the Pauli $X$ operator. Starting from $|+\rangle$ and applying
\begin{eqnarray}
R_z(\phi_i)=\exp\!\left(-i\frac{\phi_i}{2}Z\right), \qquad \phi_i=\arccos(x_i),
\end{eqnarray}
produces $\langle X\rangle = x_i$. The coefficient is thus encoded in a relative phase.

Expectation encoding interacts naturally with the phase-damping mechanism developed in Section~\ref{subsec:phase_damping}. A phase-damping channel transforms Bloch coordinates as
\begin{eqnarray}
(x_i,y_i,z_i)\mapsto(\sqrt{1-\gamma_i}\,x_i,\sqrt{1-\gamma_i}\,y_i,z_i),
\end{eqnarray}
yielding the contraction
\begin{eqnarray}
\langle X\rangle_{\text{after},i} = \sqrt{1-\gamma_i}\,x_i.
\end{eqnarray}
This is exactly the multiplicative shrinkage rule that underpins decoherence-based attenuation, and it connects directly to the controlled-dephasing. Because each coefficient resides on a separate qubit, shrinkage can be applied in parallel with scale- or location-dependent strengths.

The main limitation is that global transforms such as $U_W$ cannot be directly applied when coefficients are encoded independently. Even so, expectation encoding is experimentally simple, avoids global normalization issues, and provides a direct route to empirical calibration of shrinkage via $X$-basis readouts.

\begin{example}
To make the phase encoding construction more concrete, we consider a  signal from previous examples,
\begin{equation}
y = (2, 1, 9, 0, 3, -10, 2, 4).
\label{eq:phase_y_vector}
\end{equation}
The signal is rescaled into the interval $[-1,1]$ resulting in a vector $(d_0,\dots, d_7)$. The $d_j$ serve as the classical inputs that control the phases.

Phase encoding is implemented as a diagonal unitary on the computational basis. We first prepare a reference state on $n = 3$ qubits, for example, the uniform superposition
\begin{equation}
|\psi_0\rangle = 2^{-3/2} \sum_{j=0}^{7} |j\rangle.
\label{eq:phase_superposition}
\end{equation}
Given a scale parameter $\alpha > 0$, we associate to each coefficient $d_j$ a phase
\begin{equation}
\phi_j = \alpha d_j,
\label{eq:phase_phi_def}
\end{equation}
and define the phase encoding unitary by
\begin{equation}
U_{\phi} |j\rangle = e^{\mathrm{i}\phi_j} |j\rangle,
\qquad j = 0,\ldots,7.
\label{eq:phase_diag_unitary}
\end{equation}
In the circuit this is realized by a collection of controlled $R_z$ rotations or phase gates that implement the diagonal operator $U_{\phi}$ in Qiskit. Applying $U_{\phi}$ to $|\psi_0\rangle$ produces the encoded state
\begin{equation}
|\psi_{\mathrm{in}}\rangle 
= U_{\phi} |\psi_0\rangle
= 2^{-3/2} \sum_{j=0}^{7} e^{\mathrm{i}\phi_j} |j\rangle,
\label{eq:phase_encoded_state}
\end{equation}
which now carries the information in $y$ entirely in the phases of the computational basis components.

To model shrinkage through decoherence we pass each qubit of $|\psi_{\mathrm{in}}\rangle$ through the phase damping channel $\Phi_{\gamma}$ with Kraus operators given in (\ref{eq:kraus}). The resulting density matrix is
\begin{equation}
\rho_{\mathrm{out}} = \bigl(\Phi_{\gamma}^{\otimes 3}\bigr)\bigl(|\psi_{\mathrm{in}}\rangle\langle\psi_{\mathrm{in}}|\bigr),
\label{eq:phase_out_rho}
\end{equation}
and the attenuation of coherence appears as a multiplicative factor on the off diagonal entries of $\rho_{\mathrm{out}}$ compared with $|\psi_{\mathrm{in}}\rangle\langle\psi_{\mathrm{in}}|$.

In the notebook {\tt QWShrink12.ipynb} we summarize the effect of phase damping through simple observables. For example, we look at the expectation of a Pauli operator $X$ on the first qubit,
\begin{eqnarray}
m_{\mathrm{before}} 
&=& \langle \psi_{\mathrm{in}}| X \otimes I \otimes I |\psi_{\mathrm{in}}\rangle,
\label{eq:phase_m_before}
\\
m_{\mathrm{after}} 
&=& \mathrm{tr}\Bigl(\rho_{\mathrm{out}} \, X \otimes I \otimes I \Bigr).
\label{eq:phase_m_after}
\end{eqnarray}
For the phase damping channel these expectations satisfy a relation of the form
\begin{equation}
m_{\mathrm{after}} \approx s(\gamma) \, m_{\mathrm{before}},
\label{eq:phase_shrink_factor}
\end{equation}
where $s(\gamma)$ is a shrinkage factor that depends on the decoherence strength $\gamma$.

\begin{figure}[h!]
\centering
\includegraphics[width=0.8\textwidth]{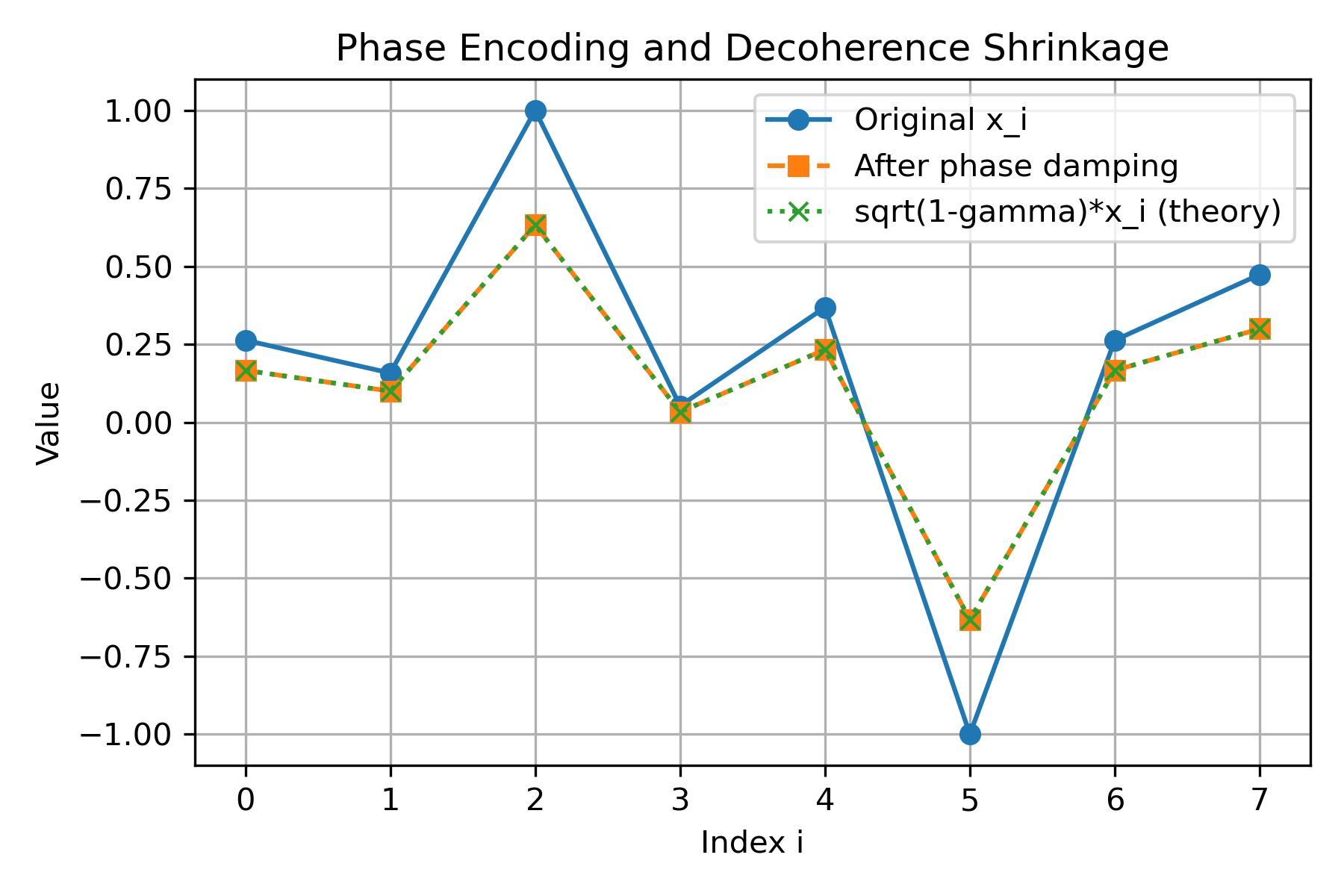} %
\caption{\small
Normalized wavelet coefficients $d_j$ obtained from the sequence
$[2, 1, 9, 0, 3, -10, 2, 4]$.  This example shows, in a very explicit and low dimensional setting, how phase encoding of a fixed signal together with phase damping implements a smooth shrinkage of phase based information.}
\label{fig:phaseencode}
\end{figure}

\end{example}

\vspace*{0.15in}
\noindent {\bf Hybrid and Mixed Encodings.}
Between amplitude and expectation encoding lies a family of hybrid schemes that allow wavelet transforms and shrinkage channels to be implemented in complementary domains. One approach applies the wavelet transform in amplitude encoding and then transfers coefficients into expectation encoding before shrinkage. If $U_W|x\rangle=|d\rangle=\sum_j d_j|j\rangle$, one prepares states $|\psi_j\rangle$ satisfying $\langle X\rangle_{\psi_j}=d_j$ and applies attenuation independently across qubits. This hybrid interface is especially useful in NISQ-era protocols where global unitaries are feasible but CPTP shrinkage is easier to apply locally.

Another hybrid representation encodes magnitude and sign separately: the magnitude $|x_j|$ is stored as an amplitude, while the sign is stored as a relative phase,
\begin{eqnarray}
|x\rangle=\frac{1}{\|x\|}\sum_j |x_j|\big(|0\rangle + e^{i\pi(1-\mathrm{sgn}(x_j))}|1\rangle\big)|j\rangle.
\end{eqnarray}
Amplitude damping then regulates magnitudes while dephasing stabilizes or regularizes sign fluctuations. This representation aligns naturally with Bayesian or SPoM-type shrinkage rules, where both magnitude and sign uncertainty influence the amount of attenuation.

\vspace*{0.15in}

Amplitude encoding supports global unitary transforms and ancilla-driven CPTP shrinkage; expectation encoding aligns with localized decoherence-based attenuation; hybrid schemes combine the advantages of both. The encoding strategy should therefore be matched to the desired level of coherence preservation and the quantum resources available. For a broader perspective on encoding choices and noise-aware simulation, see \citet{YuanEndoZhao2022}. Together, these approaches form a coherent interface between classical signal representation, physical quantum channels, and statistical inference in quantum wavelet shrinkage.

\section{Hardware Realizability and NISQ Constraints}
\label{sec:hardware}

The shrinkage mechanisms developed in Sections~\ref{subsec:ancilla_cptp}--\ref{subsec:hybrid_shrinkage} can be simulated directly in \texttt{Qiskit 2.x} or \texttt{Aer}, where Kraus operators for channels such as phase damping or amplitude damping may be applied programmatically. On real hardware, however, arbitrary Kraus maps are not native operations (as discussed with Nick Broon, from IBM). Present devices provide only unitary gates, measurements, and qubit resets. Even so, two experimentally accessible techniques permit practical realizations of shrinkage dynamics on present noisy intermediate-scale quantum platforms.

\vspace*{0.15in}
\noindent {\bf Controlled Dephasing via Calibrated Idling.}
Superconducting qubits undergo natural dephasing during idle periods, with characteristic time $T_2$. This process is equivalent to a phase-damping channel with parameter
\begin{eqnarray}
\gamma = 1 - e^{-2t/T_2},
\end{eqnarray}
where $t$ is the idle duration. To attenuate $\langle X\rangle$ by a factor $s\in[0,1]$, one idles for
\begin{eqnarray}
t \approx -T_2 \ln(s),
\end{eqnarray}
or, written in terms of $\sqrt{1-\gamma}$,
\begin{eqnarray}
t \approx -T_2 \ln \sqrt{1-\gamma}.
\end{eqnarray}
This converts a desired shrinkage level into a precisely calibrated waiting time. As described in Section~\ref{subsec:phase_damping}, the resulting attenuation reproduces the rule $\langle X_i\rangle_{\text{after}}=\sqrt{1-\gamma_i}\,\langle X_i\rangle_{\text{before}}$. By assigning different idle times to different qubits, one achieves coefficient-specific shrinkage without extra circuit depth.

\vspace*{0.15in}
\noindent {\bf Randomized Pauli-$Z$ Flips as Stochastic Dephasing.}
A digital surrogate for phase damping applies a Pauli-$Z$ gate with probability $\gamma$ and the identity with probability $1-\gamma$. Averaging measurement results over many repetitions produces the ensemble channel
\begin{eqnarray}
\mathcal{E}_\gamma(\rho) = (1-\gamma)\rho + \gamma Z\rho Z,
\end{eqnarray}
which contracts off-diagonal elements identically to physical dephasing \citep{nielsen2010}. This method requires no hardware modification; a classical random bit determines whether a $Z$ gate is applied in each shot. Ensemble averaging yields the expected contraction of $\langle X\rangle$, giving a software-level implementation of controlled shrinkage.

\vspace*{0.15in}

Calibrated idling and randomized $Z$ flips therefore provide hardware-native implementations of the shrinkage operations developed in Sections~\ref{sec:quantum_shrinkage_framework}--\ref{sec:encoding}. The first uses actual decoherence dynamics to tune attenuation continuously; the second achieves the same effect statistically by varying the probability of $Z$ flips. Both methods require only timing control, conditional gate application, and repeated sampling, all of which are native to current superconducting and trapped-ion architectures. In this light, noise becomes not merely a limitation but a functional resource. By shaping decoherence rates or flip probabilities, one can sculpt shrinkage behavior directly in hardware, transforming physical noise processes into computational tools for multiscale inference.

\section{Conclusions}
\label{sec:conclusion}

This work introduces a unified framework for quantum wavelet shrinkage, extending the classical idea of coefficient thresholding into the quantum domain through physically realizable operations.  The central contribution is conceptual: shrinkage, long viewed as a nonlinear postprocessing step in classical wavelet denoising, can be reformulated as a completely positive trace-preserving (CPTP) map implemented by controlled decoherence.  The equivalence between phase damping and multiplicative attenuation establishes a direct bridge between statistical inference and open quantum dynamics.

At the methodological level, ancilla-driven channels and phase-damping surrogates provide coherent, unitary mechanisms for attenuation of wavelet coefficients without measurement.  These constructions demonstrate that statistical adaptivity and quantum unitarity are not incompatible but can coexist within a properly designed ancilla framework.  The link between shrinkage factors and physical decoherence parameters further enables translation of statistical tuning into experimental control.

From the practical standpoint, the paper identifies two hardware level realizations, calibrated idling and randomized Pauli-$Z$ flips, that make the proposed quantum shrinkage schemes executable on current noisy intermediate-scale quantum (NISQ) devices.  Both realize the attenuation rule $\langle X_i\rangle_{\text{after}}=\sqrt{1-\gamma_i}\,\langle X_i\rangle_{\text{before}}$ using only native operations such as timing control and probabilistic gate insertion.  In this setting, decoherence ceases to be a limitation and becomes a programmable computational resource.

While the present framework realizes shrinkage through continuous CPTP attenuation, one can also view denoising as a quantum decision problem.  In that setting, significant and noise-dominated coefficients correspond to two quantum hypotheses, and a Helstrom measurement provides the minimum-error rule for distinguishing them \citep{Helstrom1976}.  Such measurement-based strategies may complement channel-driven shrinkage by offering an optimal, discrete postprocessing stage once the state has evolved through a damping map, thereby linking quantum detection theory and wavelet-based inference.

 All examples in this paper come from the annotated and executable Jupyter notebooks from the repository (\url{https://github.com/BraniV/QuantumWaveletShrinkage}), enabling reproducible experimentation and adaptation to simulators and real hardware backends.

\medskip

\noindent{\bf Acknowledgments.}
The author gratefully acknowledges support from the National Science Foundation under Grant No.~2515246 at Texas~A\&M~University and from the H.~O.~Hartley Chair research funds at Texas~A\&M.  The author also thanks Nick~Broon of IBM~Quantum for valuable discussions during his visit to Texas~A\&M, which helped shape several perspectives on the interplay between wavelet shrinkage and quantum hardware challenges.

\bibliography{references}

@inproceedings{Klappenecker1999,
  author    = {Andreas Klappenecker},
  title     = {Wavelets and Wavelet Packets on Quantum Computers},
  booktitle = {Proceedings of the 3rd International Conference on Computational Intelligence and Multimedia Applications (ICCIMA)},
  year      = {1999},
  pages     = {169--173},
  publisher = {IEEE},
  doi       = {10.1109/ICCIMA.1999.795210}
}

@article{LiLi2018,
  author    = {Zhi-Hui Li and Jian Li and Feng-Guang Li and Shou-Jun Xu},
  title     = {The multi-level and multi-dimensional quantum wavelet transform},
  journal   = {Quantum Information Processing},
  year      = {2018},
  volume    = {17},
  number    = {9},
  pages     = {240},
  doi       = {10.1007/s11128-018-2000-y}
}

@article{Bagherimehrab2023,
  author    = {Mohammad Bagherimehrab and Al{\'a}n Aspuru-Guzik},
  title     = {Efficient Quantum Algorithm for All Quantum Wavelet Transforms},
  journal   = {Quantum},
  year      = {2023},
  volume    = {7},
  pages     = {1103},
  doi       = {10.22331/q-2023-10-10-1103}
}

@article{donoho1994,
  author  = {Donoho, David L. and Johnstone, Iain M.},
  title   = {Ideal Spatial Adaptation by Wavelet Shrinkage},
  journal = {Biometrika},
  year    = {1994},
  volume  = {81},
  number  = {3},
  pages   = {425--455}
}

@article{donoho1995,
  author  = {Donoho, David L. and Johnstone, Iain M.},
  title   = {Adapting to Unknown Smoothness via Wavelet Shrinkage},
  journal = {Journal of the American Statistical Association},
  year    = {1995},
  volume  = {90},
  number  = {432},
  pages   = {1200--1224}
}

@book{vidakovic1999,
  author    = {Vidakovic, Brani},
  title     = {Statistical Modeling by Wavelets},
  publisher = {Wiley},
  year      = {1999}
}

@article{Cai1999,
  author  = {Cai, T. Tony},
  title   = {Adaptive Wavelet Estimation: A Block Thresholding and Oracle Inequality Approach},
  journal = {Annals of Statistics},
  year    = {1999},
  volume  = {27},
  number  = {3},
  pages   = {898--924}
}

@article{LiWang2018,
  author  = {Li, H.-S. and Wang, S.-C. and others},
  title   = {Multi-level and Multi-dimensional Quantum Wavelet Packet Transform},
  journal = {Quantum Information Processing},
  year    = {2018},
  volume  = {17},
  number  = {6},
  pages   = {133}
}

@article{li2019,
  author  = {Li, H.-S. and others},
  title   = {Quantum Wavelet Transforms for General Orthogonal Wavelets},
  journal = {Quantum Information Processing},
  year    = {2019},
  volume  = {18},
  number  = {2},
  pages   = {50}
}

@misc{nakahira2021,
  author       = {Nakahira, Kenta},
  title        = {Efficient Implementation of Quantum Orthogonal Wavelet Transforms},
  howpublished = {Preprint},
  year         = {2021},
  eprint       = {arXiv:2107.09016}
}

@article{kraus1971,
  author  = {Kraus, Karl},
  title   = {General State Changes in Quantum Theory},
  journal = {Annals of Physics},
  year    = {1971},
  volume  = {64},
  number  = {2},
  pages   = {311--335}
}

@book{nielsen2010,
  author    = {Nielsen, Michael A. and Chuang, Isaac L.},
  title     = {Quantum Computation and Quantum Information},
  publisher = {Cambridge University Press},
  year      = {2010}
}

@book{lidar2013,
  author    = {Lidar, Daniel A. and Brun, Todd A.},
  title     = {Quantum Error Correction},
  publisher = {Cambridge University Press},
  year      = {2013}
}

@article{givens1958,
  author    = {Wallace Givens},
  title     = {Computation of Plane Unitary Rotations Transforming a General Matrix to Triangular Form},
  journal   = {Journal of the Society for Industrial and Applied Mathematics},
  year      = {1958},
  volume    = {6},
  number    = {1},
  pages     = {26--50},
  doi       = {10.1137/0106003}
}

@inproceedings{hoyer1997,
  author    = {Peter Høyer},
  title     = {Efficient Quantum Transforms},
  booktitle = {Proceedings of the 24th International Colloquium on Automata, Languages and Programming (ICALP)},
  year      = {1997},
  pages     = {144--155},
  publisher = {Springer},
  address   = {Berlin, Heidelberg},
  doi       = {10.1007/3-540-63165-8_184}
}

@inproceedings{fijany1998,
  author    = {A. Fijany and C. P. Williams},
  title     = {Quantum Wavelet Transforms: Fast Algorithms and Complete Circuits},
  booktitle = {Proceedings of the SPIE Conference on Quantum Computing},
  year      = {1998},
  volume    = {3076},
  pages     = {175--183},
  doi       = {10.1117/12.302230}
}

@inproceedings{fijany1998springer,
  author    = {A. Fijany and C. P. Williams},
  title     = {Quantum Wavelet Transforms and Their Implementation},
  booktitle = {Quantum Computing and Quantum Communications},
  editor    = {Colin P. Williams},
  series    = {Lecture Notes in Computer Science},
  volume    = {1509},
  pages     = {10--33},
  publisher = {Springer},
  address   = {Berlin, Heidelberg},
  year      = {1998},
  doi       = {10.1007/BFb0053328}
}

@article{aharonov1988,
  author  = {Yakir Aharonov and David Z. Albert and Lev Vaidman},
  title   = {How the result of a measurement of a component of the spin of a spin-$\tfrac{1}{2}$ particle can turn out to be 100},
  journal = {Physical Review Letters},
  year    = {1988},
  volume  = {60},
  number  = {14},
  pages   = {1351--1354},
  doi     = {10.1103/PhysRevLett.60.1351}
}

@book{jacobs2014,
  author    = {Kurt Jacobs},
  title     = {Quantum Measurement Theory and its Applications},
  publisher = {Cambridge University Press},
  year      = {2014}
}

@book{wiseman2009,
  author    = {H. M. Wiseman and Gerard J. Milburn},
  title     = {Quantum Measurement and Control},
  publisher = {Cambridge University Press},
  year      = {2009}
}

@article{stinespring1955,
  author       = {Stinespring, W. Forrest},
  title        = {Positive functions on C*-algebras},
  journal      = {Proceedings of the American Mathematical Society},
  volume       = {6},
  number       = {2},
  pages        = {211--216},
  year         = {1955},
  doi          = {10.2307/2032342}
}

@article{choi1975,
  author       = {Choi, Man-Duen},
  title        = {Completely positive linear maps on complex matrices},
  journal      = {Linear Algebra and Its Applications},
  volume       = {10},
  number       = {3},
  pages        = {285--290},
  year         = {1975},
  doi          = {10.1016/0024-3795(75)90075-0}
}

@article{lindblad1976,
  author       = {Lindblad, G\"oran},
  title        = {On the generators of quantum dynamical semigroups},
  journal      = {Communications in Mathematical Physics},
  volume       = {48},
  number       = {2},
  pages        = {119--130},
  year         = {1976},
  doi          = {10.1007/BF01608499}
}

@article{GarciaMata2009,
  author  = {Garc{\'i}a-Mata, Ignacio and Giraud, Olivier and Georgeot, Bertrand},
  title   = {Quantum computation of multifractal exponents through the quantum wavelet transform},
  journal = {Physical Review A},
  year    = {2009},
  volume  = {79},
  number  = {6},
  pages   = {062324},
  doi     = {10.1103/PhysRevA.79.062324}
}

@article{Ma2024,
  author  = {Ma, Guang and Li, Jia and Duan, Zhen and Xu, Kai and Zhang, Hongwei},
  title   = {Great-length wavelets on quantum computing platform},
  journal = {Signal Processing},
  year    = {2024},
  volume  = {216},
  pages   = {109365},
  doi     = {10.1016/j.sigpro.2024.109365}
}

@article{ChaurraGutierrez2023,
  author  = {Chaurra Guti{\'e}rrez, F. A. and Mej{\'i}a-Lavalle, M. and V{\'a}squez-Medina, R. and Mu{\~n}oz, R.},
  title   = {One-dimensional Quantum Integer CDF(2,2) Wavelet Transform},
  journal = {arXiv preprint},
  year    = {2023},
  eprint  = {arXiv:2312.02874}
}

@article{Zhang2019,
  author  = {Zhang, Xiangyu and Wang, Yong and Zhang, Li and Xu, Shoujun},
  title   = {Dimension Reduction Using Quantum Wavelet Transform},
  journal = {Journal of Advanced Computational Intelligence and Intelligent Informatics},
  year    = {2019},
  volume  = {23},
  number  = {2},
  pages   = {194--202},
  doi     = {10.20965/jaciii.2019.p0194}
}

@book{SchuldPetruccione2018,
  author    = {Schuld, Maria and Petruccione, Francesco},
  title     = {Supervised Learning with Quantum Computers},
  publisher = {Springer},
  address   = {Cham},
  year      = {2018},
  series    = {Quantum Science and Technology},
  doi       = {10.1007/978-3-319-96424-9}
}

@article{BiamonteEtAl2017,
  author    = {Biamonte, Jacob and Wittek, Peter and Pancotti, Nicola and Rebentrost, Patrick and Wiebe, Nathan and Lloyd, Seth},
  title     = {Quantum Machine Learning},
  journal   = {Nature},
  year      = {2017},
  volume    = {549},
  number    = {7671},
  pages     = {195--202},
  doi       = {10.1038/nature23474}
}

@article{YuanEndoZhao2022,
  author    = {Yuan, Xiao and Endo, Suguru and Zhao, Qisheng and Li, Ying and Benjamin, Simon C.},
  title     = {Theory of Variational Quantum Simulation},
  journal   = {Reviews of Modern Physics},
  year      = {2022},
  volume    = {94},
  number    = {1},
  pages     = {015004},
  doi       = {10.1103/RevModPhys.94.015004}
}

@book{Helstrom1976,
  author    = {Helstrom, Carl W.},
  title     = {Quantum Detection and Estimation Theory},
  publisher = {Academic Press},
  address   = {New York},
  year      = {1976}
}

@article{VidakovicRuggeri2001,
  author  = {Brani Vidakovic and Fabrizio Ruggeri},
  title   = {{BAMS} Method: Theory and Simulations},
  journal = {Sankhyā, Series B},
  year    = {2001},
  volume  = {63},
  number  = {2},
  pages   = {234--249},
  note    = {Special issue on wavelets},
  doi     = {10.1007/s13171-005-0004-x}
}

@article{Nason1996,
  author    = {Nason, Guy P.},
  title     = {Wavelet Shrinkage Using Cross-Validation},
  journal   = {Journal of the American Statistical Association},
  volume    = {91},
  number    = {434},
  pages     = {1206--1220},
  year      = {1996},
  doi       = {10.1080/01621459.1996.10476967}
}

@article{Vidakovic1998,
  author    = {Vidakovic, Brani},
  title     = {Nonlinear Wavelet Shrinkage with {B}ayes Rules and {B}ayes Factors},
  journal   = {Journal of the American Statistical Association},
  volume    = {93},
  number    = {441},
  pages     = {173--179},
  year      = {1998},
  doi       = {10.1080/01621459.1998.10473757}
}

@article{Chipman1997,
  author    = {Chipman, Hugh A. and Kolaczyk, Eric D. and McCulloch, Robert E.},
  title     = {Adaptive {B}ayesian Wavelet Shrinkage},
  journal   = {Journal of the American Statistical Association},
  volume    = {92},
  number    = {440},
  pages     = {1413--1421},
  year      = {1997},
  doi       = {10.1080/01621459.1997.10474053}
}

@techreport{GosalLawton2001,
  author    = {Gosal, Darwin and Lawton, Wayne},
  title     = {Quantum Haar Wavelet Transforms and Their Applications},
  institution = {National University of Singapore},
  address   = {Singapore},
  year      = {2001},
  note      = {Technical Report, November 5, 2001},
  url       = {https://web.cecs.pdx.edu/~mperkows/CAPSTONES/HAAR/quantum12.pdf}
}

@inproceedings{PuschelMouraFijany2001,
  author    = {Püschel, Markus and Moura, José M. F. and Fijany, Amir},
  title     = {Quantum Algorithms for Wavelet Transforms on Quantum Computers},
  booktitle = {Proceedings of the IEEE International Conference on Acoustics, Speech, and Signal Processing (ICASSP)},
  year      = {2001},
  pages     = {2129--2132},
  doi       = {10.1109/ICASSP.2001.940613}
}

@inproceedings{DasDattaFijany2002,
  author    = {Das, Sudebkumar and Datta, Animesh and Fijany, Amir},
  title     = {Quantum Multiresolution Analysis and Quantum Wavelet Transforms},
  booktitle = {Proceedings of the 2002 International Symposium on Circuits and Systems (ISCAS)},
  year      = {2002},
  pages     = {IV-337--IV-340},
  doi       = {10.1109/ISCAS.2002.1011181}
}

@article{Geller2024,
  author    = {Geller, Michael R.},
  title     = {Protocol for Nonlinear State Discrimination in Rotating Condensate},
  journal   = {Advanced Quantum Technologies},
  volume    = {7},
  number    = {4},
  year      = {2024},
  doi       = {10.1002/qute.202300431},
  note      = {Nonlinear positive trace-preserving dynamics for enhanced state discrimination}
}
\end{document}